\documentclass[%
 reprint,
 amsmath,amssymb,
 aps,
]{revtex4-2}

\usepackage{mathtools}
\usepackage{amsfonts}
\usepackage{graphicx}
\usepackage{dcolumn}
\usepackage{bm}
\usepackage{booktabs} 
\usepackage{arydshln} 
\usepackage{makecell}
\usepackage[justification=raggedright,singlelinecheck=false]{caption}
\usepackage{subfig}  

\begin{document}

\preprint{APS/123-QED}

\title{The impact of digital media-driven affective polarisation on epidemic dynamics}

\author{Satoshi Komuro}
\email{\href{mailto:satoshi.komuro@auckland.ac.nz}{satoshi.komuro@auckland.ac.nz}}
\

\begin{abstract}
While prior studies have examined the influence of information diffusion on epidemic dynamics, the role of affective polarisation—driven by digital media usage—remains less understood. This study introduces a mathematical framework to quantify the interplay between affective polarisation and epidemic spread, revealing contrasting effects depending on transmission rates. The model demonstrates that greater digital media influence leads to increased polarisation. Notably, the results reveal opposing trends: a negative correlation between polarisation and the infected population is observed when transmission rates are low, whereas a positive correlation emerges in high-transmission scenarios. These findings provide a quantitative foundation for assessing how digital media-driven polarisation may exacerbate health crises, informing future public health strategies.
\end{abstract}

\maketitle


\section{Introduction}
\label{sec:introduction}

The interplay between information diffusion and epidemic dynamics has been widely studied, with models capturing how health-related behaviours spread through populations \cite{funk2010modelling, verelst2016behavioural, manfredi2013modeling}. However, these studies often overlook the role of polarisation—a process wherein individuals align strongly with like-minded peers while dismissing opposing views, particularly in digital media environments. Given the increasing political and social polarisation observed in public health debates, understanding its impact on disease spread is essential.

Throughout history, humanity has contended with a wide range of infectious diseases, including smallpox, the plague, and the Spanish flu \cite{mcneill1976plagues}. More recently, the COVID-19 pandemic has caused substantial loss of life and widespread societal disruption. A key factor influencing the spread of infectious diseases is individual behaviour \cite{funk2010modelling}. The extent to which populations adopt preventive measures, such as vaccination, mask-wearing, lockdowns, and social distancing, can significantly alter epidemic dynamics. In recognition of this, various mathematical models have been developed to capture the spread of infectious diseases while accounting for the collective behavioural responses of populations \cite{verelst2016behavioural, manfredi2013modeling, funk2009spread}.

With the rise of digital media, particularly social networking services (SNS), people are now able to access, gather, spread, and create information instantly. Consequently, both the quantity and quality of information circulating in society have evolved, influencing \textit{opinion dynamics}. During the COVID-19 pandemic, digital platforms became flooded with information, both accurate and inaccurate. Some of this content is considered \textit{misinformation} (false or misleading information shared without harmful intent) \cite{wardle2017information}, while other content is classified as \textit{disinformation} (false or misleading information shared with harmful intent) \cite{fallis2015disinformation}. The sheer volume of information spread at an unprecedented rate, mirroring the rapid transmission of an infectious disease. This phenomenon was widely recognised as an \textit{``infodemic,”} a term that garnered significant attention \cite{zarocostas2020fight}. As WHO Director-General Tedros Adhanom Ghebreyesus stated at the 56th Munich Security Conference: ``We’re not just fighting an epidemic; we’re fighting an infodemic. Fake news spreads faster and more easily than this virus, and is just as dangerous” \cite{WHO_Munich2025_Infodemic}.

The term ``infodemic" and the phrase ``go viral," commonly used to describe rapidly spreading content online, highlight the striking similarities between information dissemination in digital media and epidemic spread. Mathematically, these processes can be modelled in similar ways, leading to growing research interest in the coupled dynamics of information diffusion and epidemic transmission. Many studies in this area apply \textit{compartmental models}, originally developed for epidemic modelling, to information diffusion as well, facilitating parallel modelling for analytical simplicity. However, while these models capture some aspects of information spread, they often fail to account for the role of \textit{polarisation}, a key social phenomenon observed in public health debates.

During the COVID-19 pandemic, \textit{societal polarisation} emerged between vaccine proponents and opponents \cite{henkel2023association}. Additionally, \textit{affective polarisation}, where individuals show affinity toward those with similar views and hostility toward those with opposing views, played a critical role \cite{filsinger2024asymmetric}. Political identity has also been linked to vaccine attitudes: in the United States, vaccine hesitancy correlated with support for Donald Trump \cite{hornsey2020donald}, whereas in Japan, it was more closely associated with liberalism \cite{toriumi2024anti}. In Europe, research on conspiracy theorists found no direct link between vaccine scepticism and political orientation but identified correlations with educational background and a negative association between religiosity and education level \cite{jabkowski2023exploring}. These findings suggest that opinions on vaccination extend beyond mere health concerns, serving as markers of broader political and social identities.

This raises an important question: How does polarisation, often amplified by digital media, influence epidemic dynamics? To address this, we build on the framework proposed by T\"{o}rnberg (2022) \cite{tornberg2022digital}, which demonstrates that when \textit{homophily} (the tendency to interact with like-minded individuals) is strong, affective polarisation tends to intensify as the probability of engaging in random interactions, akin to exchanges on digital media, increases. These results suggest that greater digital media influence drives affective polarisation.

In this paper, we investigate how digital media-driven polarisation influences epidemic dynamics across different scenarios. Specifically, we model the interplay between information diffusion and epidemic spread on multiplex networks, adopting T\"{o}rnberg’s information diffusion model for the information layer and a compartmental model for the epidemic layer. Our findings provide insights into how digital media can shape public health outcomes, with potential implications for policy interventions aimed at mitigating the effects of misinformation and polarisation.

\section{Related Work}
\label{sec:related_work}

This section reviews previous work on mathematical models designed to capture information diffusion, epidemic spreading, and the interactions between these processes. 

\subsection{Compartmental Model}
A foundational mathematical framework for modelling such processes is the compartmental model, which categorises individuals in a population into distinct states (compartments) and describes transitions between these states. One of the earliest and most influential models in this context is the SIR (Susceptible-Infected-Recovered) model \cite{kermack1927contribution}. Proposed in the 1920s, the SIR model was designed to represent the spread of infectious diseases. Its simplicity and effectiveness have led to its widespread application in epidemiology and beyond. In this model, individuals are classified into three compartments: Susceptible (S), Infected (I), and Recovered (R). The transitions between these states are governed by the rates of infection and recovery, providing a framework for understanding the dynamics of disease spread within a population. Infection spreads when susceptible individuals come into contact with infected individuals, with a transition rate proportional to the infection rate $\beta$. Infected individuals recover at a constant rate, represented by the recovery rate $\gamma$. The dynamics of these transitions are defined by the following system of ordinary differential equations:
\begin{subequations}\label{det_SIR_eqs}
\begin{align}
&\frac{dS}{dt} = -\beta SI, \label{SIR_ds}\\
&\frac{dI}{dt} = \beta SI - \gamma I, \label{SIR_di} \\
&\frac{dR}{dt} = \gamma I. \label{SIR_dr}
\end{align}
\end{subequations}

Although this paper does not delve into the detailed analysis of these equations, analysing them provides insights into key epidemiological properties, such as the peak number of infected individuals and the final number of recovered individuals. In particular, setting $S(0) = S_0$, it is well known that the spread and containment of the disease are determined by the threshold $\mathcal{R}_0 \coloneqq \beta S_0 / \gamma$. This threshold, known as the \textit{basic reproduction number}, indicates that if $\mathcal{R}_0$ is greater than 1, the infection spreads, while if it is 1 or lower, it subsides. The basic reproduction number is a crucial metric for controlling and managing disease outbreaks. Moreover, various compartmental models have been proposed to accommodate different epidemiological characteristics, such as the SIS (Susceptible-Infected-Susceptible) model, which assumes that recovered individuals become susceptible again, and the SEIR (Susceptible-Exposed-Infected-Recovered) model, which introduces an exposed (E) compartment to account for an incubation period before infection.

Furthermore, by transforming Equations \ref{det_SIR_eqs} into a continuous-time Markov chain:
\begin{subequations}\label{stochastic_SIR}
\begin{align}
& (S, I, R) \xrightarrow{\beta S I}(S-1, I+1, R), \\
& (S, I, R) \xrightarrow{\gamma I}(S, I-1, R+1),
\end{align}
\end{subequations}
we obtain a stochastic model \cite{kiss2017mathematics}. Since disease spread inherently involves randomness, particularly in small populations or early-stage outbreaks where individual interactions significantly impact overall dynamics, incorporating stochasticity is crucial.

One approach to modelling disease transmission more realistically is to introduce network structures \cite{keeling2005networks, pastor2015epidemic}. In the network-based SIR model, the infection rate for a susceptible node linked to $k$ infected nodes is given by $1 - (1 - \beta)^k$. Advances in network science \cite{barabasi2016network, newman2018networks} have revealed that many real-world systems can be modelled as networks with shared structural properties. Specifically, \textit{scale-free networks}, where node degrees follow a power-law distribution, exhibit significant heterogeneity among nodes. Traditional models assume that agents interact randomly with any other agent, an unrealistic assumption. In contrast, network-based models account for local interaction structures and heterogeneity.

Compartmental models are not limited to epidemiology; they have been applied to various diffusion processes, such as opinion dynamics \cite{zhao2013sir}. In this context, individuals are classified by opinion states, with transitions representing opinion changes. For instance, an SIR-like model can be adapted for opinion spread by assigning $S$ to neutral or uninfluenced individuals, $I$ to those who have adopted a particular opinion and actively spread it, and $R$ to those who have dropped out of the opinion-spreading process or become resistant to new opinions. The transmission rate ($\beta$) represents opinion spread, while the recovery rate ($\gamma$) represents the rate at which individuals stop expressing or change their opinions. Such models have been used to analyse phenomena like \textit{echo chambers} and polarisation \cite{cinelli2021echo}.

\subsection{UAU-SIS Model}
Efforts have been made to unify social and epidemiological phenomena by modelling information diffusion and disease dissemination as coevolving spreading processes on multiplex networks \cite{granell2013dynamical, granell2014competing, guo2015two, guo2016epidemic, peng2021multilayer, PhysRevE.102.022312, xia2019new, fang2023coevolution}. A pioneering study by Granell et al. \cite{granell2013dynamical} assumes that each agent has both an infection state (either $S$ for Susceptible or $I$ for Infected) and an awareness state (either $U$ for Unaware or $A$ for Aware). Disease and information spread are modelled using SIS processes on separate network layers. Since the information layer consists of $U$ and $A$ states, this model is called the UAU-SIS model. Here, $\beta_A = \epsilon \beta_U$ ($0 \leq \epsilon \leq 1$) ensures that aware agents have a lower infection probability than unaware agents. Moreover, infected individuals automatically transition to the aware state. By employing the Microscopic Markov Chain Approach (MMCA), the model’s steady states can be mathematically analysed, revealing that the epidemic threshold depends on the information layer’s dynamics.

However, real-world applications face challenges. The model oversimplifies information diffusion and does not fully capture real-world opinion dynamics. Specifically, assuming symmetry between disease spread (which requires physical contact) and information spread (which can transcend geographic and temporal constraints via the internet) is a strong assumption requiring careful consideration. Thus, this study primarily contributes to the theoretical understanding of multilayer network dynamics rather than real-world applications.

\subsection{Polarisation Model through Partisan Sorting}
One pressing issue in modern society is polarisation and societal division. Empirical and theoretical research has sought to understand this phenomenon. Although the echo chamber hypothesis is widely discussed as a driver of polarisation, some empirical studies do not support it \cite{dubois2018echo}. This highlights the need for further theoretical investigation into polarisation mechanisms.

\textit{Partisan sorting}, where individuals group based on political or social affiliations and homogenise opinions within these groups \cite{mason2015disrespectfully}, is a known phenomenon. T\"ornberg (2022) proposed that partisan sorting, rather than echo chambers, drives affective polarisation in digital media. Using an extension of Axelrod’s cultural dissemination model \cite{axelrod1997dissemination}, agents interact randomly and adopt attributes from others. However, interactions occur not only between neighbouring agents but also over long distances in a network. Agents preferentially interact with similar others (homophily), reflecting how digital media enables interaction regardless of physical distance.

Simulations on different networks show that when homophily surpasses a critical threshold, increased long-distance interactions accelerate polarisation. This simple model effectively captures modern information diffusion and reproduces polarisation, making it a suitable framework for studying the relationship between polarisation and disease dynamics.

\section{Model Overview}
\label{sec:model}

\begin{figure}
    \centering
    \includegraphics[width=\linewidth]{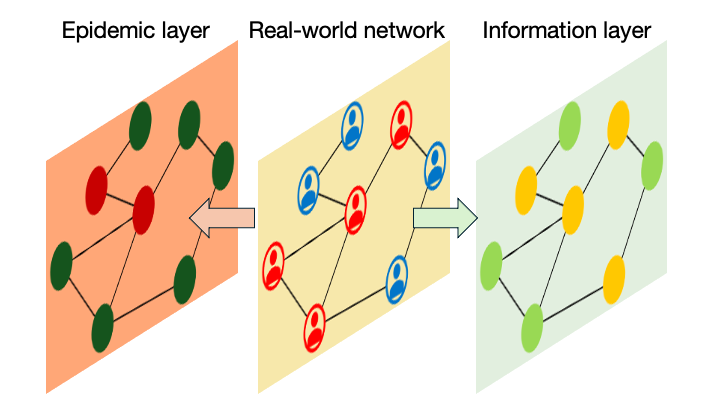}
    \caption{A schematic representation of the proposed model at a given time step. The central part of the figure represents the network that simulates real-world social interactions, where each node belongs to one of two partisan groups, denoted by red and blue colours. The left panel illustrates the epidemic layer, where each node is in one of two states: $S$ (Susceptible) or $I$ (Infected), represented by red and green nodes, respectively. The right panel depicts the information layer, where each node holds opinions on five topics. The yellow and light green colours represent awareness states regarding the epidemic: $U$ (Unaware) and $A$ (Aware).}
    \label{fig:model_description}
\end{figure}

In this study, we represent the network of social interactions as a graph $G = (V, E)$. The set of nodes $V$ represents individuals, and in our simulations, we set $N \coloneqq |V| = 1000$. The set of edges $E \subseteq (V \times V)$ represents social ties, generated using the Holme-Kim model \cite{holme2002growing} to ensure a scale-free degree distribution. This model extends the Barab\'asi-Albert model \cite{albert2002statistical} by incorporating preferential attachment for new nodes and a probability $p = 0.01$ of forming additional links between neighbouring nodes, enhancing local clustering.

Each node $i \in V$ is assigned a fixed partisan affiliation $S_i \in \{1, ..., k\}$ that remains unchanged throughout the simulation and is not influenced by interactions. In our simulations, we set $k = 2$. 

Additionally, each node $i$ has a state vector $W_i(t) = (D_i(t), E_i(t))$, where $D_i(t)$ represents the opinion vector in the information layer and $E_i(t)$ represents the infection state in the epidemic layer. These states evolve according to different mechanisms on the same network topology. In this study, the network topology remains static, and only the state vectors of the nodes change over time. Unless otherwise stated, state updates occur in both the information and epidemic layers at each time step, and we record the system’s state after a fixed number of steps. By varying key parameters, we conduct multiple simulations to examine the interplay between digital media, polarisation, and epidemic spread. Figure \ref{fig:model_description} provides a schematic of the model at a given time step.

\subsection{Information layer}

In the information layer, each node $i$ possesses a dynamic opinion vector of length $n$, denoted as $D_i \in \{1, ..., m\}^{n-1} \times \{U, A\}$. In our simulations, we set $m = 3$ and $n = 5$. This configuration represents opinions on four topics, each with three possible stances (modelling agreement, disagreement, and neutrality), along with an awareness state regarding the epidemic: $U$ (Unaware) or $A$ (Aware). Initially, the opinion vectors of all nodes are assigned randomly.

At each time step, one node $i$ is randomly selected for an opinion update. Let $\Gamma_i$ be the set of its neighbours and $k_i \coloneqq |\Gamma_i|$ be the degree of node $i$. From $\Gamma_i$, we randomly select $\lfloor (1 - \gamma) k_i \rfloor$ nodes, and from the entire network $V$, we randomly select $\lceil \gamma k_i \rceil$ nodes to form the interaction set $\mathcal{I}$. The parameter $\gamma \in [0, 1]$ controls the influence of digital media: a larger $\gamma$ increases the proportion of nodes in $\mathcal{I}$ that are selected from $V$ rather than $\Gamma_i$. 

A node $j \in \mathcal{I}$ is chosen for interaction based on the following probabilities:

\begin{align}
  \delta_{ij}^S &\coloneqq 
  \begin{cases} 
    1 & \text{if } S_i = S_j, \\ 
    0 & \text{otherwise}. \label{eq:fix_similarity}
  \end{cases} \\
  \delta_{ij, l}^D &\coloneqq
  \begin{cases} 
    1 & \text{if } D_{i} \cdot \mathbf{e}_l = D_{j} \cdot \mathbf{e}_l, \\ 
    0 & \text{otherwise}. \label{eq:flex_similarity}
  \end{cases}
\end{align}

where $\mathbf{e}_l$ is a unit basis vector for $1 \leq l \leq n$. 

The absolute similarity $\delta_{ij}$ between nodes $i$ and $j$ is defined as:

\begin{equation}
\delta_{ij} \coloneqq \frac{c \delta_{ij}^S + \sum\limits_{l=1}^{n} \delta_{ij, l}^D}{c + n}.
\end{equation}

Here, $c$ is a weighting factor for partisanship; a larger $c$ gives greater importance to partisan alignment over opinion similarity. We fix $c = 2$ in our simulations.

The probability that node $j$ is chosen for interaction is given by:

\begin{equation}
p_{ij} \coloneqq \frac{{\delta_{ij}}^h}{\sum\limits_{k \in \mathcal{I}} {\delta_{ik}}^h }.
\end{equation}

The parameter $h$ determines homophily, with larger values increasing the likelihood that $i$ interacts with more similar nodes. We fix $h = 32$ in our simulations.

Once node $j$ is selected, if $\sum_{l=1}^{n} \delta_{ij, l}^D \neq n$, one differing opinion component $l$ is randomly chosen, and $D_i$'s $l$th element is updated to match $D_j$. If all components already match ($\sum_{l=1}^{n} \delta_{ij, l}^D = n$), no update occurs.

\subsection{Epidemic layer}

In the epidemic layer, we model disease spread using the SIS (Susceptible-Infected-Susceptible) model. Each node \( i \) has an infection state \( E_i \in \{S, I\} \), where \( S \) denotes a susceptible state and \( I \) represents an infected state. The model evolves according to the following parameters:

\begin{itemize}
    \item Infection rate \( \beta \)
    \item Recovery rate \( \mu \)
    \item Initial infection probability \( \rho^I_0 = 0.2 \)
\end{itemize}

At each time step, a node is selected randomly, and its state is updated based on its interactions with neighbours. The transition probabilities are given by:

\begin{align}
  \mathbb{P}(E_i(t+1)=I \mid E_i(t)=S) &= 1-(1-\beta)^{k_i^{I}}, \label{eq:infection_prob} \\
  \mathbb{P}(E_i(t+1)=S \mid E_i(t)=I) &= \mu. \label{eq:recovery_prob}
\end{align}

where \( k_i^I \) denotes the number of infected neighbours of node \( i \). Equation \eqref{eq:infection_prob} represents the probability that at least one infected neighbour transmits the disease to node \( i \) during a time step.

\subsection{Inter-layer interactions}

The interaction between the information and epidemic layers is modelled through awareness and infection risk. Each node has a dynamic opinion vector in the information layer, where one component represents awareness of the disease.

Awareness affects infection risk by reducing the probability of infection. Following Granell et al. \cite{granell2013dynamical}, aware nodes have a lower probability of infection compared to unaware nodes. Specifically, the infection rate for aware individuals is given by:

\begin{equation}
    \beta_A = \epsilon \beta_U, \quad 0 \leq \epsilon \leq 1.
\end{equation}

When \( \epsilon = 0 \), aware individuals do not get infected, whereas when \( \epsilon = 1 \), awareness has no effect on infection probability.

Conversely, the epidemic layer influences the information layer by resetting awareness. A node that recovers from the infected state loses its awareness and becomes unaware again. This models situations where individuals become less cautious over time after recovering from an illness.

These interactions between awareness and disease spread capture the co-evolution of information diffusion and epidemic dynamics within a polarised social network.

\section{Results}
\label{sec:results}
The results of the simulation based on the model described in Section \ref{sec:model} are presented in this section. The parameter $\gamma$ is varied from 0 to 1 in increments of 0.02, while other parameters are set to their default values as listed in Table \ref{tab:parameters}. For each value of $\gamma$, the simulation results after 50,000 steps are recorded 100 times, and the results are analysed.

\begin{table}[ht!]
\centering
\resizebox{\columnwidth}{!}{
\begin{tabular}{ccc}
\toprule
\textbf{Parameter} & \textbf{Description} & \textbf{Default Value} \\
\midrule
$N$ & Number of nodes & 1000 \\
$k$ & Number of parties & 2 \\
\Xhline{1pt}
\addlinespace[2pt]  
$n$ & Opinion dimension & 5 \\
$m$ & Number of opinions per topic & 3 \\
$\gamma$ & Digital media influence & $[0, 1]$ \\
$c$ & Partisanship weight & 2 \\
$h$ & Homophily & 32 \\
$\psi$ & Level of polarisation & - \\
$\rho^A_0$ & Initial aware proportion & 0.5\\
$\rho^A$ & Aware population proportion & - \\
\Xhline{1pt}
\addlinespace[2pt]  
$\beta$ & Infection rate & 0.05 \\
$\mu$ & Recovery rate & 0.01 \\
$\rho^I_{0}$ & Initial infection proportion &  0.2 \\
$\rho^I$ & Infected population proportion & - \\
\bottomrule
\end{tabular}
}
\caption{List of parameters used in the model. The top section includes network parameters that are relevant to both layers. The middle section details parameters for the information layer, while the bottom section describes parameters for the epidemic layer.}
\label{tab:parameters}
\end{table}

The level of polarisation $\psi$ is defined as follows, following Törnberg (2022). First, for any nodes $i, j \in V$, the opinion agreement $d_{ij}$ is calculated using $\delta_{ij, l}^D$ from equation \eqref{eq:flex_similarity}:

\begin{equation}
d_{ij} \coloneqq \dfrac{\sum\limits_{i=1}^{n} \delta_{ij, l}^D}{n} 
\end{equation}

where 
\begin{align} 
Q_{\text{same}} &\coloneqq {d_{ij} \mid i, j \in V, i \neq j, S_i = S_j }, \\
\ Q_{\text{diff}} &\coloneqq {d_{ij} \mid i, j \in V, i \neq j, S_i \neq S_j }.
\end{align}

$Q_{\text{same}}$ is the set of opinion agreements between nodes belonging to the same party, and $Q_{\text{diff}}$ is the set of opinion agreements between nodes belonging to different parties. Next, the level of polarisation $\psi$ is expressed as:

\begin{equation} 
\psi \coloneqq \frac{\sum_{d_{ij} \in Q_{\text{same}}} d_{ij}}{|Q_{\text{same}}|} - \frac{\sum_{d_{ij} \in Q_{\text{diff}}} d_{ij}}{|Q_{\text{diff}}|}. 
\end{equation}

In a fully polarised state, when all nodes belonging to the same party have identical opinion vectors and the opinion vectors of nodes from different parties share no common elements, $\psi = 1$. On the other hand, when the opinion vectors of nodes are independent of partisanship, $\mathbb{E}(\psi) = 0$.

Furthermore, the proportion of the aware population $\rho^A$ and the proportion of the infected population $\rho^I$ are defined as follows: 
\begin{align}
\rho^A &\coloneqq \dfrac{1}{N} \left| { i \in V \mid D_{i} \cdot \mathbf{e}_n = A } \right|, \\
\ \rho^I &\coloneqq \dfrac{1}{N} \left| { i \in V \mid E_i = I } \right|. 
\end{align}

\subsection{Digital media and its relationship to affective polarisation}
Törnberg (2022) exhibited that polarisation progresses as the influence of digital media increases, under the assumption that \( h \) is sufficiently high. However, the model used in the simulation couples the infection layers, introducing feedback through inter-layer interactions, resulting in fundamentally different dynamics from the setting of Törnberg (2022). Therefore, it is not self-evident whether the relationship between the influence of digital media \( \gamma \) and the level of polarisation \( \psi \) will yield the same results as in Törnberg (2022).

\begin{figure}[htbp!] 
    \centering 
    \includegraphics[width=\linewidth]{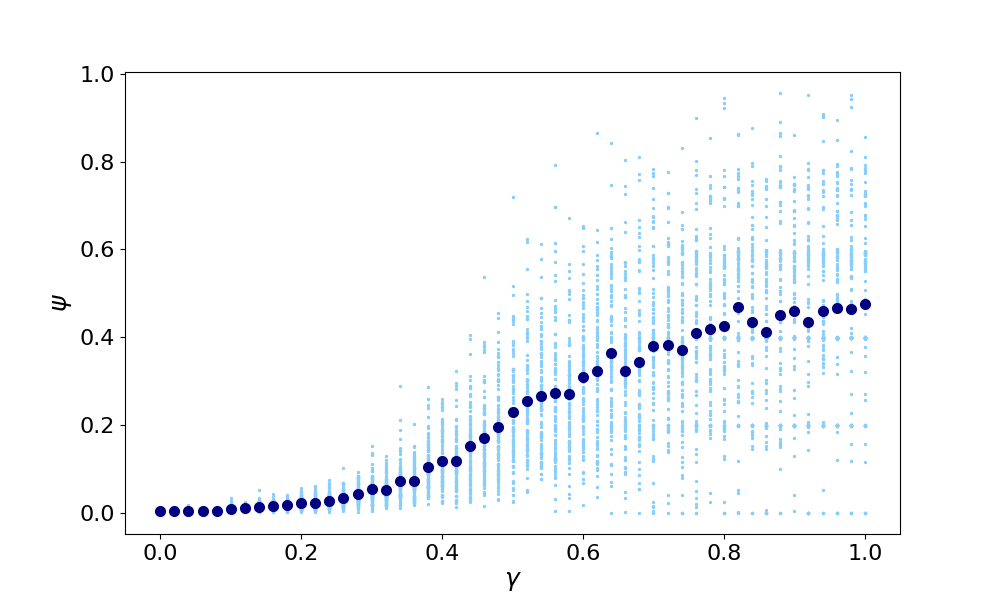} 
    \caption{The relationship between the probability $\gamma$ of interacting with nodes other than adjacent nodes during information updates and the level of polarisation $\psi$. $\gamma$ is varied from 0 to 1 in increments of 0.02, resulting in 51 patterns. For each value of $\gamma$, 100 simulations are performed, and the $\psi$ values after 50,000 steps are plotted in light blue, with the average value plotted in dark blue. As $\gamma$ increases, $\psi$ also increases. All other parameter values follow Table \ref{tab:parameters}.} 
    \label{fig:scatter} 
\end{figure}

The relationship between $\gamma$ and $\psi$ is shown in Figure \ref{fig:scatter}. The value of $\gamma$ is taken along the horizontal axis, and the final values of $\psi$ are plotted in light blue. Additionally, the mean of $\psi$ for each value of $\gamma$ is calculated and plotted in dark blue. This result indicates that, similar to Törnberg's (2022) study, in this model, once a certain threshold is exceeded, an increase in $\gamma$ tends to lead to an increase in the level of polarisation $\psi$.

\subsection{Affective polarisation and its relationship to infected population}
As outlined in the previous part, it was presented that as the influence of digital media increases, polarisation progresses in the model used in this study. Next, we investigate the relationship between the level of polarisation $\psi$ and the proportion of infected individuals $\rho^I$, which is the central issue of this paper.

\begin{figure}[htbp!]
    \centering
    \subfloat[\centering Scatter plot of $\rho^I$ ($\beta = 0.05$, $\mu = 0.01$)]{
    \centering
    \includegraphics[width=0.45\linewidth]{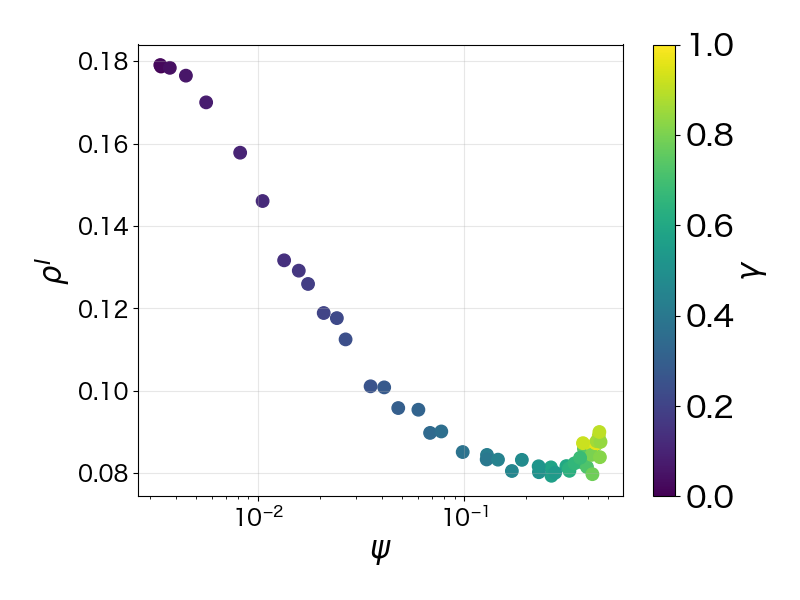}
    \label{fig:low_rho_I}
    }
    \subfloat[\centering Scatter plot of $\rho^I$ ($\beta = 0.005$, $\mu = 0.01$)]{
        \centering
        \includegraphics[width=0.45\linewidth]{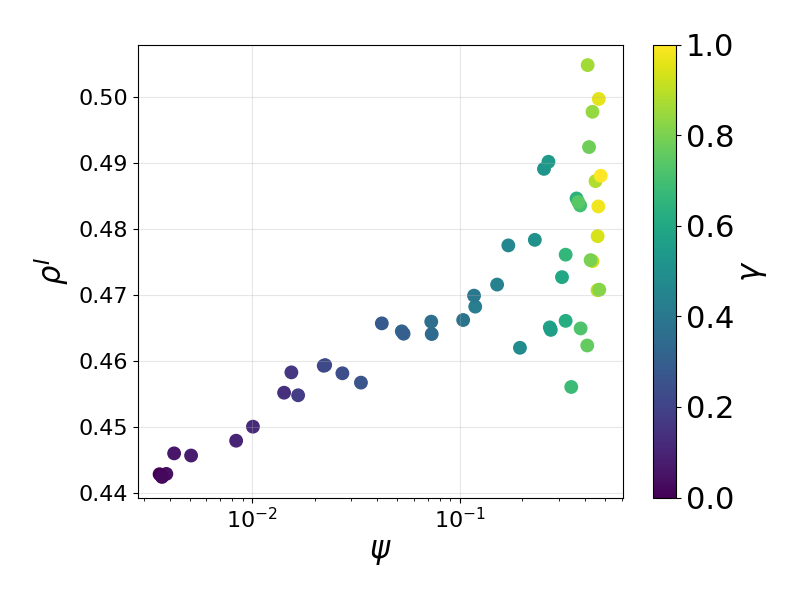}
        \label{fig:high_rho_I}
    }
    \quad
    \caption{Comparison of scatter plots of the level of polarisation \(\psi\) and the proportion of infected individuals \(\rho^I\). The colour differences represent the digital media influence \(\gamma\). As in Figure \ref{fig:scatter}, for each value of \(\gamma\), the \(\psi\) values after 50000 steps are calculated, and the average over 100 simulations is plotted. In Figure \ref{fig:low_rho_I}, a downward trend is observed, while in Figure \ref{fig:high_rho_I}, an upward trend is obtained.}
    \label{fig:comparison}
\end{figure}

In this section, we consider two scenarios: one in which the infection situation is mild and one in which it is severe. Two different combinations of $\beta$ and $\mu$ were adopted, and the results of analysing the relationship between the level of polarisation $\psi$ and the proportion of infected individuals $\rho^I$ are shown in Figure \ref{fig:comparison}. The upper panel shows the results for $\beta = 0.005$, $\mu = 0.1$, and the lower panel shows the results for $\beta = 0.05$, $\mu = 0.01$. In the latter simulation, the infection state was updated once every 10 iterations of information updates.

Interestingly, contrasting results were obtained in these two scenarios. The correlation coefficient between the logarithms of $\psi$ and $\rho^I$ was $r = -0.923$ in Figure \ref{fig:low_rho_I} and $r = 0.816$ in Figure \ref{fig:high_rho_I}. That is, in the scenario with mild infection, there was a strong negative correlation between $\psi$ and $\rho^I$, whereas in the scenario with severe infection, a strong positive correlation was observed.

\subsection{Heatmap analysis on infected population}
In the preceding section, we illustrated that the impact of polarisation on the spread of infection varies depending on whether the infection situation is mild or severe. Based on this result, a natural point of interest is how \(\rho^I\) evolves in response to changes in \(\beta\).  

\begin{figure}
    \centering
    \includegraphics[width=\linewidth]{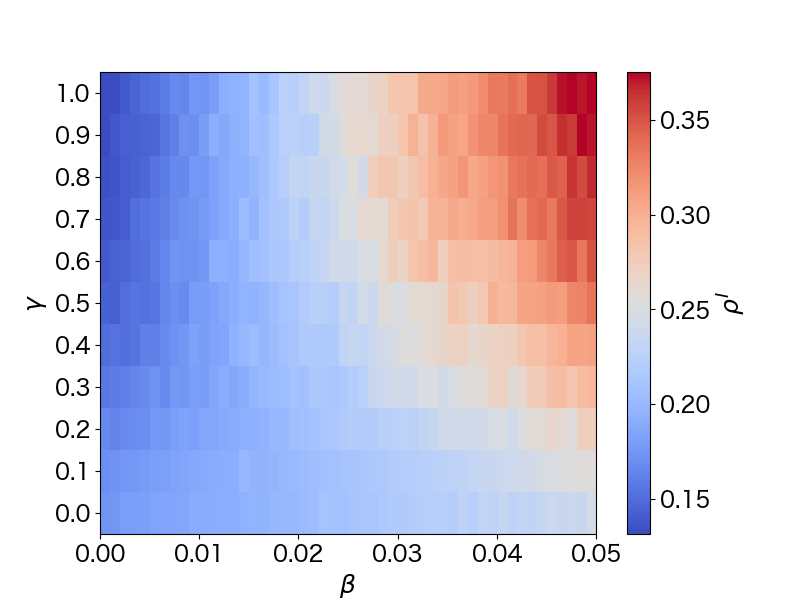}
    \caption{Heatmap of \(\rho^I\) when \(\beta\) and \(\gamma\) are varied. \(\beta\) was varied in 50 steps from 0.001 to 0.05 in increments of 0.001, and \(\gamma\) was varied in 11 steps from 0 to 1 in increments of 0.1, resulting in 550 combinations. For each combination, 10 simulations were run, and the average \(\rho^I\) after 5000 steps was plotted. \(\mu = 0.1\) and other parameters follow those in Table \ref{tab:parameters}.}
    \label{fig:heatmap}
\end{figure}

Figure \ref{fig:heatmap} presents a heatmap illustrating the distribution of the proportion of infected individuals \(\rho^I\) as \(\beta\) and \(\gamma\) vary. Higher values of \(\rho^I\) are represented by red shades, while lower values are shown in blue. As already observed in Figure \ref{fig:comparison}, when \(\beta\) is small, \(\rho^I\) decreases as \(\gamma\) increases. In contrast, when \(\beta\) is large, \(\rho^I\) increases with \(\gamma\). In the intermediate region, the colours transition smoothly, indicating that \(\rho^I\) changes gradually with respect to \(\beta\) rather than undergoing an abrupt shift at a specific threshold. Additionally, it can be inferred that the larger the value of \(\gamma\), the more sensitively \(\rho^I\) responds to changes in \(\beta\), whereas smaller values of \(\gamma\) result in a weaker influence.

\section{Discussion}
As highlighted in Section \ref{sec:results}, the model shows that once a certain threshold of $\gamma$ is exceeded, an increase in $\gamma$ leads to a rise in the level of polarisation $\psi$. This suggests that higher levels of digital media influence tend to increase emotional polarisation within the network. Regarding the relationship between polarisation $\psi$ and the proportion of the infected population $\rho^I$, the results indicate polar-opposite trends depending on the infection rate $\beta$ and the recovery rate $\mu$. Specifically, when $\beta$ is small and $\mu$ is large (mimicking situations where the infection is mild), a negative correlation is observed between polarisation $\psi$ and $\rho^I$. In contrast, when $\beta$ is large and $\mu$ is small (representing more severe infection scenarios), a positive correlation emerges between these two variables.

\begin{figure}[htbp!]
    \centering
    \subfloat[\centering Scatter plot of $\rho^A$ ($\beta = 0.005$, $\mu = 0.1$)]{
        \includegraphics[width=0.45\linewidth]{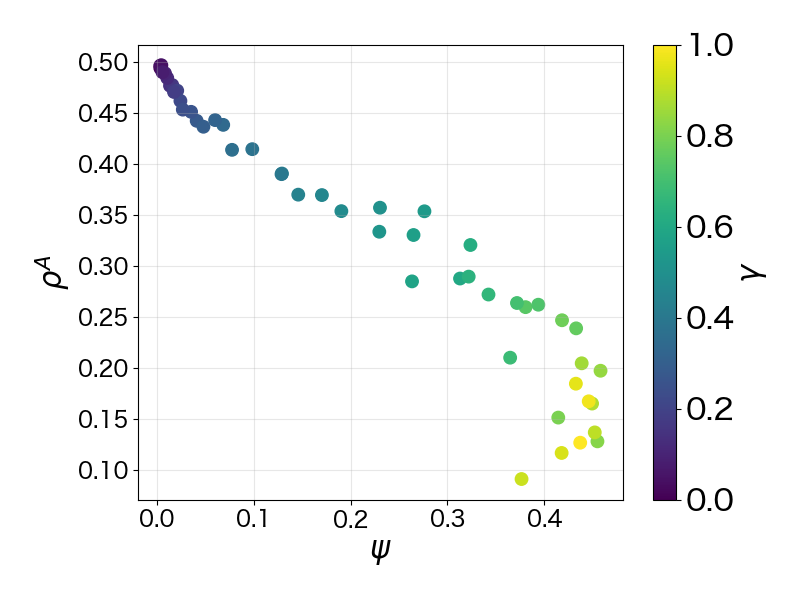}
        \label{fig:low_rho_A}
    }%
    \hfill
    \subfloat[\centering Residual scatter plot of $\rho^I$ ($\beta = 0.005$, $\mu = 0.1$)]{
        \includegraphics[width=0.45\linewidth]{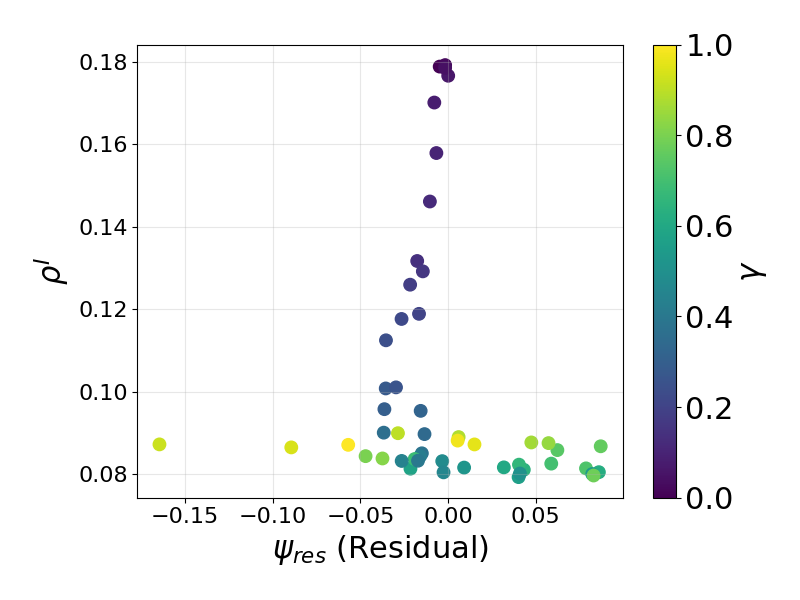}
        \label{fig:low_rho_I_residuals}
    }
    \vspace{0.1em} 
    \subfloat[\centering Scatter plot of $\rho^A$ ($\beta = 0.05$, $\mu = 0.01$)]{
        \includegraphics[width=0.45\linewidth]{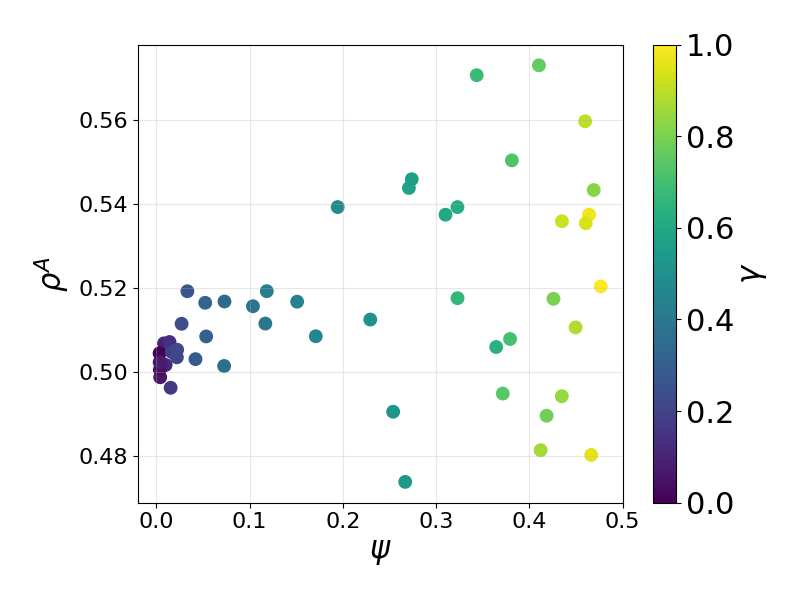}
        \label{fig:high_rho_A}
    }%
    \hfill
    \subfloat[\centering Residual scatter plot of $\rho^I$ ($\beta = 0.05$, $\mu = 0.01$)]{
        \includegraphics[width=0.45\linewidth]{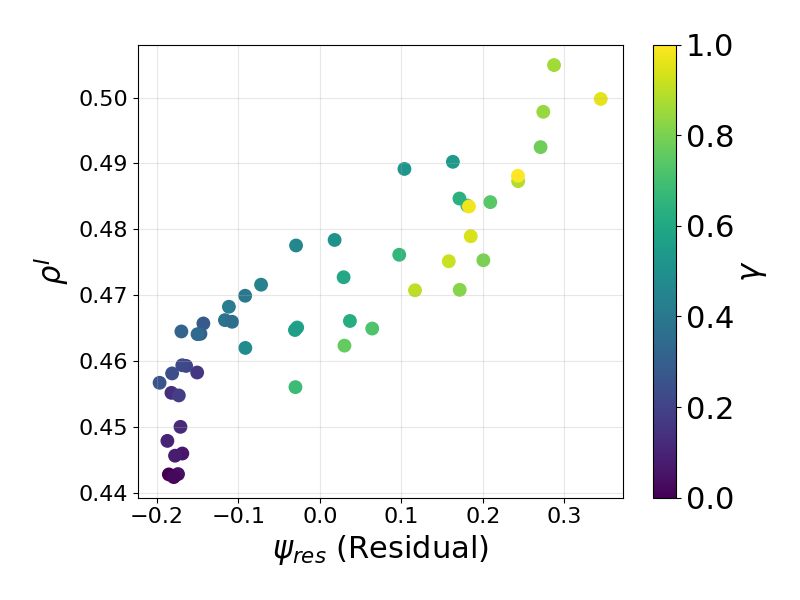}
        \label{fig:high_rho_I_residuals}
    }
    \caption{Scatter plots of the level of polarisation \(\psi\) versus the proportion of aware individuals \(\rho^A\), along with scatter plots of \(\rho^I\) plotted against the residual \(\psi_{\text{res}}\). The top row corresponds to the case of \(\beta = 0.005\), \(\mu = 0.1\), and the bottom row to \(\beta = 0.05\), \(\mu = 0.01\). Each data point represents the average over 100 simulations for each \(\gamma\), calculated after 50000 steps. The other parameter values follow Table \ref{tab:parameters}.}  
\label{fig:scatter_aware_res}
\end{figure}

A useful insight into understanding this phenomenon comes from the relationship between polarisation $\psi$ and the aware population proportion $\rho^A$. As shown in Figure \ref{fig:low_rho_A}, when the infection rate is low, a strong negative correlation of $r = -0.963$ is observed between $\psi$ and $\rho^A$. Statistically, by removing the influence of $\psi$ on $\rho^A$, the residuals and their relationship with $\rho^I$ are shown in Figure \ref{fig:low_rho_I_residuals}, with a correlation of $r = -0.161$. This suggests that the relationship between $\psi$ and $\rho^I$ is likely a pseudo-correlation induced by the awareness population $\rho^A$, and that polarisation itself is not directly contributing to infection prevention.

Conversely, in the case of a higher infection rate, as seen in Figure \ref{fig:high_rho_A}, the correlation between $\psi$ and $\rho^A$ is weaker, with a value of $r = 0.343$. Similarly, after removing the influence of $\psi$ on $\rho^A$, the residual scatter plot in Figure \ref{fig:high_rho_I_residuals} shows a strong positive correlation of $r = 0.886$ with $\rho^I$. This suggests that, in this scenario, polarisation itself is more strongly correlated with the proportion of the infected population.

In this model, partisanship is initially assigned randomly and does not change over time, nor is it influenced by interactions. Therefore, in a model with only two partisan groups, as polarisation progresses, both the aware and unaware populations are randomly distributed across the network, much like partisanship itself. In contrast, when polarisation is low, the aware and unaware populations tend to be clustered, particularly in local communities where the aware population may exceed a certain threshold.

This phenomenon can be likened to the concept of \textit{herd immunity} \cite{fine2011herd}. When a sufficient proportion of the population holds immunity, the spread of infection can be mitigated even among those who are not immune. In this model, we may interpret the emergence of awareness as a form of herd immunity, where polarisation results in the loss of this collective immunity, leading to an increase in the infected population.

Moreover, it is generally more crucial to consider the dynamics during the later stages of an epidemic rather than in the early, milder stages. The findings from this model suggest that during a pandemic, the active use of digital media could accelerate emotional polarisation, which may, in turn, exacerbate the spread of the infection. This underscores the importance of understanding the potential negative feedback loops between polarisation and disease spread in crisis situations, particularly when emotional and social divides may hinder coordinated responses.

\section{Conclusion and future work}
This study models the spread of information and infectious diseases as diffusion processes occurring on different layers of multiplex networks to analyse how affective polarisation induced by digital media influences the dynamics of disease transmission. In this model, digital media enables interactions with like-minded individuals regardless of geographical proximity. As a result, it was observed that the level of polarisation increases as the influence of digital media intensifies.  

Furthermore, an analysis of the relationship between polarisation and the infected population revealed a negative correlation when the infection rate was low, whereas a positive correlation was observed when the infection rate was high. This indicates that the direction of the impact of polarisation on disease spread varies depending on the strength of the infection rate.  

When the infection rate was low, a strong negative correlation was found between polarisation and the population aware of the disease. However, after removing the influence of the disease-aware population, no correlation was found between polarisation and the infected population, suggesting that the initial correlation was a spurious one, mediated by the disease-aware population as a latent variable. In contrast, when the infection rate was high, no strong correlation was observed between polarisation and the disease-aware population, and even after removing its influence, a strong positive correlation between polarisation and the infected population remained. This suggests that, in such cases, polarisation itself may contribute to the spread of the disease.  

In general, public interest in infection control measures tends to increase as the severity of an outbreak worsens. The findings of this study suggest that during a pandemic, active engagement with digital media accelerates affective polarisation, which in turn can exacerbate the spread of infections. While societal divisions have traditionally been discussed from the perspectives of social sciences and political economy, this study highlights their potential risks from the viewpoint of infectious disease control.

\subsection{Future work}
This work observes a negative correlation between polarisation and awareness population when the infection rate is low. However, a deeper investigation is needed to understand the underlying mechanisms. Also, the current study assumes random partisanship, which may affect the final epidemic size when polarisation occurs. In real-world settings, individuals who share partisanship may be likely to be connected in some way, meaning the assumption of randomness may not hold. Future work should explore more realistic assumptions, such as examining the relationship between community structure and final epidemic size, to obtain more nuanced insights.

Additionally, while the current results are derived from simulations, a data-driven approach is necessary to assess how well this model reflects real-world dynamics. By comparing the model’s predictions with actual data, we can gain a better understanding of its applicability and validity.

Finally, to make the model more analytically tractable, future work could focus on developing a simpler version of the model, which would allow for mathematical analysis and provide further insight into the dynamics of polarisation and epidemic spread.







\section*{Data availability}
The codes underlying this article are available on \href{https://github.com/skomuro/polarisation_meets_epi/}{the dedicated Github repository} \cite{git_repo}. The data itself has not been uploaded due to its large file size; however, following the instructions and running the code will reproduce the exact same data and figures. Additionally, the data used in this paper is available upon request.

\bibliography{main}

\providecommand{\noopsort}[1]{}\providecommand{\singleletter}[1]{#1}%
\begin{thebibliography}{38}%
\makeatletter
\providecommand \@ifxundefined [1]{%
 \@ifx{#1\undefined}
}%
\providecommand \@ifnum [1]{%
 \ifnum #1\expandafter \@firstoftwo
 \else \expandafter \@secondoftwo
 \fi
}%
\providecommand \@ifx [1]{%
 \ifx #1\expandafter \@firstoftwo
 \else \expandafter \@secondoftwo
 \fi
}%
\providecommand \natexlab [1]{#1}%
\providecommand \enquote  [1]{``#1''}%
\providecommand \bibnamefont  [1]{#1}%
\providecommand \bibfnamefont [1]{#1}%
\providecommand \citenamefont [1]{#1}%
\providecommand \href@noop [0]{\@secondoftwo}%
\providecommand \href [0]{\begingroup \@sanitize@url \@href}%
\providecommand \@href[1]{\@@startlink{#1}\@@href}%
\providecommand \@@href[1]{\endgroup#1\@@endlink}%
\providecommand \@sanitize@url [0]{\catcode `\\12\catcode `\$12\catcode `\&12\catcode `\#12\catcode `\^12\catcode `\_12\catcode `\%12\relax}%
\providecommand \@@startlink[1]{}%
\providecommand \@@endlink[0]{}%
\providecommand \url  [0]{\begingroup\@sanitize@url \@url }%
\providecommand \@url [1]{\endgroup\@href {#1}{\urlprefix }}%
\providecommand \urlprefix  [0]{URL }%
\providecommand \Eprint [0]{\href }%
\providecommand \doibase [0]{https://doi.org/}%
\providecommand \selectlanguage [0]{\@gobble}%
\providecommand \bibinfo  [0]{\@secondoftwo}%
\providecommand \bibfield  [0]{\@secondoftwo}%
\providecommand \translation [1]{[#1]}%
\providecommand \BibitemOpen [0]{}%
\providecommand \bibitemStop [0]{}%
\providecommand \bibitemNoStop [0]{.\EOS\space}%
\providecommand \EOS [0]{\spacefactor3000\relax}%
\providecommand \BibitemShut  [1]{\csname bibitem#1\endcsname}%
\let\auto@bib@innerbib\@empty
\bibitem [{\citenamefont {Funk}\ \emph {et~al.}(2010)\citenamefont {Funk}, \citenamefont {Salath{\'e}},\ and\ \citenamefont {Jansen}}]{funk2010modelling}%
  \BibitemOpen
  \bibfield  {author} {\bibinfo {author} {\bibfnamefont {S.}~\bibnamefont {Funk}}, \bibinfo {author} {\bibfnamefont {M.}~\bibnamefont {Salath{\'e}}},\ and\ \bibinfo {author} {\bibfnamefont {V.~A.}\ \bibnamefont {Jansen}},\ }\bibfield  {title} {\bibinfo {title} {Modelling the influence of human behaviour on the spread of infectious diseases: a review},\ }\href@noop {} {\bibfield  {journal} {\bibinfo  {journal} {Journal of the Royal Society Interface}\ }\textbf {\bibinfo {volume} {7}},\ \bibinfo {pages} {1247} (\bibinfo {year} {2010})}\BibitemShut {NoStop}%
\bibitem [{\citenamefont {Verelst}\ \emph {et~al.}(2016)\citenamefont {Verelst}, \citenamefont {Willem},\ and\ \citenamefont {Beutels}}]{verelst2016behavioural}%
  \BibitemOpen
  \bibfield  {author} {\bibinfo {author} {\bibfnamefont {F.}~\bibnamefont {Verelst}}, \bibinfo {author} {\bibfnamefont {L.}~\bibnamefont {Willem}},\ and\ \bibinfo {author} {\bibfnamefont {P.}~\bibnamefont {Beutels}},\ }\bibfield  {title} {\bibinfo {title} {Behavioural change models for infectious disease transmission: a systematic review (2010--2015)},\ }\href@noop {} {\bibfield  {journal} {\bibinfo  {journal} {Journal of The Royal Society Interface}\ }\textbf {\bibinfo {volume} {13}},\ \bibinfo {pages} {20160820} (\bibinfo {year} {2016})}\BibitemShut {NoStop}%
\bibitem [{\citenamefont {Manfredi}\ and\ \citenamefont {D'Onofrio}(2013)}]{manfredi2013modeling}%
  \BibitemOpen
  \bibfield  {author} {\bibinfo {author} {\bibfnamefont {P.}~\bibnamefont {Manfredi}}\ and\ \bibinfo {author} {\bibfnamefont {A.}~\bibnamefont {D'Onofrio}},\ }\href@noop {} {\emph {\bibinfo {title} {Modeling the interplay between human behavior and the spread of infectious diseases}}}\ (\bibinfo  {publisher} {Springer Science \& Business Media},\ \bibinfo {year} {2013})\BibitemShut {NoStop}%
\bibitem [{\citenamefont {McNeill}(1976)}]{mcneill1976plagues}%
  \BibitemOpen
  \bibfield  {author} {\bibinfo {author} {\bibfnamefont {W.~H.}\ \bibnamefont {McNeill}},\ }\href@noop {} {\emph {\bibinfo {title} {Plagues and Peoples}}}\ (\bibinfo  {publisher} {Anchor Press},\ \bibinfo {year} {1976})\BibitemShut {NoStop}%
\bibitem [{\citenamefont {Funk}\ \emph {et~al.}(2009)\citenamefont {Funk}, \citenamefont {Gilad}, \citenamefont {Watkins},\ and\ \citenamefont {Jansen}}]{funk2009spread}%
  \BibitemOpen
  \bibfield  {author} {\bibinfo {author} {\bibfnamefont {S.}~\bibnamefont {Funk}}, \bibinfo {author} {\bibfnamefont {E.}~\bibnamefont {Gilad}}, \bibinfo {author} {\bibfnamefont {C.}~\bibnamefont {Watkins}},\ and\ \bibinfo {author} {\bibfnamefont {V.~A.}\ \bibnamefont {Jansen}},\ }\bibfield  {title} {\bibinfo {title} {The spread of awareness and its impact on epidemic outbreaks},\ }\href@noop {} {\bibfield  {journal} {\bibinfo  {journal} {Proceedings of the National Academy of Sciences}\ }\textbf {\bibinfo {volume} {106}},\ \bibinfo {pages} {6872} (\bibinfo {year} {2009})}\BibitemShut {NoStop}%
\bibitem [{\citenamefont {Wardle}\ and\ \citenamefont {Derakhshan}(2017)}]{wardle2017information}%
  \BibitemOpen
  \bibfield  {author} {\bibinfo {author} {\bibfnamefont {C.}~\bibnamefont {Wardle}}\ and\ \bibinfo {author} {\bibfnamefont {H.}~\bibnamefont {Derakhshan}},\ }\href@noop {} {\emph {\bibinfo {title} {Information disorder: Toward an interdisciplinary framework for research and policymaking}}},\ Vol.~\bibinfo {volume} {27}\ (\bibinfo  {publisher} {Council of Europe Strasbourg},\ \bibinfo {year} {2017})\BibitemShut {NoStop}%
\bibitem [{\citenamefont {Fallis}(2015)}]{fallis2015disinformation}%
  \BibitemOpen
  \bibfield  {author} {\bibinfo {author} {\bibfnamefont {D.}~\bibnamefont {Fallis}},\ }\bibfield  {title} {\bibinfo {title} {What is disinformation?},\ }\href@noop {} {\bibfield  {journal} {\bibinfo  {journal} {Library trends}\ }\textbf {\bibinfo {volume} {63}},\ \bibinfo {pages} {401} (\bibinfo {year} {2015})}\BibitemShut {NoStop}%
\bibitem [{\citenamefont {Zarocostas}(2020)}]{zarocostas2020fight}%
  \BibitemOpen
  \bibfield  {author} {\bibinfo {author} {\bibfnamefont {J.}~\bibnamefont {Zarocostas}},\ }\bibfield  {title} {\bibinfo {title} {How to fight an infodemic},\ }\href@noop {} {\bibfield  {journal} {\bibinfo  {journal} {The lancet}\ }\textbf {\bibinfo {volume} {395}},\ \bibinfo {pages} {676} (\bibinfo {year} {2020})}\BibitemShut {NoStop}%
\bibitem [{\citenamefont {Ghebreyesus}(2025)}]{WHO_Munich2025_Infodemic}%
  \BibitemOpen
  \bibfield  {author} {\bibinfo {author} {\bibfnamefont {T.~A.}\ \bibnamefont {Ghebreyesus}},\ }\href@noop {} {\bibinfo {title} {Speech at the munich security conference}},\ \bibinfo {howpublished} {\url{https://www.who.int/director-general/speeches/detail/munich-security-conference}} (\bibinfo {year} {2025}),\ \bibinfo {note} {accessed: 7 March 2025}\BibitemShut {NoStop}%
\bibitem [{\citenamefont {Henkel}\ \emph {et~al.}(2023)\citenamefont {Henkel}, \citenamefont {Sprengholz}, \citenamefont {Korn}, \citenamefont {Betsch},\ and\ \citenamefont {B{\"o}hm}}]{henkel2023association}%
  \BibitemOpen
  \bibfield  {author} {\bibinfo {author} {\bibfnamefont {L.}~\bibnamefont {Henkel}}, \bibinfo {author} {\bibfnamefont {P.}~\bibnamefont {Sprengholz}}, \bibinfo {author} {\bibfnamefont {L.}~\bibnamefont {Korn}}, \bibinfo {author} {\bibfnamefont {C.}~\bibnamefont {Betsch}},\ and\ \bibinfo {author} {\bibfnamefont {R.}~\bibnamefont {B{\"o}hm}},\ }\bibfield  {title} {\bibinfo {title} {The association between vaccination status identification and societal polarization},\ }\href@noop {} {\bibfield  {journal} {\bibinfo  {journal} {Nature Human Behaviour}\ }\textbf {\bibinfo {volume} {7}},\ \bibinfo {pages} {231} (\bibinfo {year} {2023})}\BibitemShut {NoStop}%
\bibitem [{\citenamefont {Filsinger}\ and\ \citenamefont {Freitag}(2024)}]{filsinger2024asymmetric}%
  \BibitemOpen
  \bibfield  {author} {\bibinfo {author} {\bibfnamefont {M.}~\bibnamefont {Filsinger}}\ and\ \bibinfo {author} {\bibfnamefont {M.}~\bibnamefont {Freitag}},\ }\bibfield  {title} {\bibinfo {title} {Asymmetric affective polarization regarding covid-19 vaccination in six european countries},\ }\href@noop {} {\bibfield  {journal} {\bibinfo  {journal} {Scientific Reports}\ }\textbf {\bibinfo {volume} {14}},\ \bibinfo {pages} {15919} (\bibinfo {year} {2024})}\BibitemShut {NoStop}%
\bibitem [{\citenamefont {Hornsey}\ \emph {et~al.}(2020)\citenamefont {Hornsey}, \citenamefont {Finlayson}, \citenamefont {Chatwood},\ and\ \citenamefont {Begeny}}]{hornsey2020donald}%
  \BibitemOpen
  \bibfield  {author} {\bibinfo {author} {\bibfnamefont {M.~J.}\ \bibnamefont {Hornsey}}, \bibinfo {author} {\bibfnamefont {M.}~\bibnamefont {Finlayson}}, \bibinfo {author} {\bibfnamefont {G.}~\bibnamefont {Chatwood}},\ and\ \bibinfo {author} {\bibfnamefont {C.~T.}\ \bibnamefont {Begeny}},\ }\bibfield  {title} {\bibinfo {title} {Donald trump and vaccination: The effect of political identity, conspiracist ideation and presidential tweets on vaccine hesitancy},\ }\href@noop {} {\bibfield  {journal} {\bibinfo  {journal} {Journal of Experimental Social Psychology}\ }\textbf {\bibinfo {volume} {88}},\ \bibinfo {pages} {103947} (\bibinfo {year} {2020})}\BibitemShut {NoStop}%
\bibitem [{\citenamefont {Toriumi}\ \emph {et~al.}(2024)\citenamefont {Toriumi}, \citenamefont {Sakaki}, \citenamefont {Kobayashi},\ and\ \citenamefont {Yoshida}}]{toriumi2024anti}%
  \BibitemOpen
  \bibfield  {author} {\bibinfo {author} {\bibfnamefont {F.}~\bibnamefont {Toriumi}}, \bibinfo {author} {\bibfnamefont {T.}~\bibnamefont {Sakaki}}, \bibinfo {author} {\bibfnamefont {T.}~\bibnamefont {Kobayashi}},\ and\ \bibinfo {author} {\bibfnamefont {M.}~\bibnamefont {Yoshida}},\ }\bibfield  {title} {\bibinfo {title} {Anti-vaccine rabbit hole leads to political representation: the case of twitter in japan},\ }\href@noop {} {\bibfield  {journal} {\bibinfo  {journal} {Journal of Computational Social Science}\ ,\ \bibinfo {pages} {1}} (\bibinfo {year} {2024})}\BibitemShut {NoStop}%
\bibitem [{\citenamefont {Jabkowski}\ \emph {et~al.}(2023)\citenamefont {Jabkowski}, \citenamefont {Domaradzki},\ and\ \citenamefont {Baranowski}}]{jabkowski2023exploring}%
  \BibitemOpen
  \bibfield  {author} {\bibinfo {author} {\bibfnamefont {P.}~\bibnamefont {Jabkowski}}, \bibinfo {author} {\bibfnamefont {J.}~\bibnamefont {Domaradzki}},\ and\ \bibinfo {author} {\bibfnamefont {M.}~\bibnamefont {Baranowski}},\ }\bibfield  {title} {\bibinfo {title} {Exploring covid-19 conspiracy theories: education, religiosity, trust in scientists, and political orientation in 26 european countries},\ }\href@noop {} {\bibfield  {journal} {\bibinfo  {journal} {Scientific Reports}\ }\textbf {\bibinfo {volume} {13}},\ \bibinfo {pages} {18116} (\bibinfo {year} {2023})}\BibitemShut {NoStop}%
\bibitem [{\citenamefont {T{\"o}rnberg}(2022)}]{tornberg2022digital}%
  \BibitemOpen
  \bibfield  {author} {\bibinfo {author} {\bibfnamefont {P.}~\bibnamefont {T{\"o}rnberg}},\ }\bibfield  {title} {\bibinfo {title} {How digital media drive affective polarization through partisan sorting},\ }\href@noop {} {\bibfield  {journal} {\bibinfo  {journal} {Proceedings of the National Academy of Sciences}\ }\textbf {\bibinfo {volume} {119}},\ \bibinfo {pages} {e2207159119} (\bibinfo {year} {2022})}\BibitemShut {NoStop}%
\bibitem [{\citenamefont {Kermack}\ and\ \citenamefont {McKendrick}(1927)}]{kermack1927contribution}%
  \BibitemOpen
  \bibfield  {author} {\bibinfo {author} {\bibfnamefont {W.~O.}\ \bibnamefont {Kermack}}\ and\ \bibinfo {author} {\bibfnamefont {A.~G.}\ \bibnamefont {McKendrick}},\ }\bibfield  {title} {\bibinfo {title} {A contribution to the mathematical theory of epidemics},\ }\href@noop {} {\bibfield  {journal} {\bibinfo  {journal} {Proceedings of the royal society of london. Series A, Containing papers of a mathematical and physical character}\ }\textbf {\bibinfo {volume} {115}},\ \bibinfo {pages} {700} (\bibinfo {year} {1927})}\BibitemShut {NoStop}%
\bibitem [{\citenamefont {Kiss}\ \emph {et~al.}(2017)\citenamefont {Kiss}, \citenamefont {Miller}, \citenamefont {Simon} \emph {et~al.}}]{kiss2017mathematics}%
  \BibitemOpen
  \bibfield  {author} {\bibinfo {author} {\bibfnamefont {I.~Z.}\ \bibnamefont {Kiss}}, \bibinfo {author} {\bibfnamefont {J.~C.}\ \bibnamefont {Miller}}, \bibinfo {author} {\bibfnamefont {P.~L.}\ \bibnamefont {Simon}}, \emph {et~al.},\ }\bibfield  {title} {\bibinfo {title} {Mathematics of epidemics on networks},\ }\href@noop {} {\bibfield  {journal} {\bibinfo  {journal} {Cham: Springer}\ }\textbf {\bibinfo {volume} {598}},\ \bibinfo {pages} {31} (\bibinfo {year} {2017})}\BibitemShut {NoStop}%
\bibitem [{\citenamefont {Keeling}\ and\ \citenamefont {Eames}(2005)}]{keeling2005networks}%
  \BibitemOpen
  \bibfield  {author} {\bibinfo {author} {\bibfnamefont {M.~J.}\ \bibnamefont {Keeling}}\ and\ \bibinfo {author} {\bibfnamefont {K.~T.}\ \bibnamefont {Eames}},\ }\bibfield  {title} {\bibinfo {title} {Networks and epidemic models},\ }\href@noop {} {\bibfield  {journal} {\bibinfo  {journal} {Journal of the royal society interface}\ }\textbf {\bibinfo {volume} {2}},\ \bibinfo {pages} {295} (\bibinfo {year} {2005})}\BibitemShut {NoStop}%
\bibitem [{\citenamefont {Pastor-Satorras}\ \emph {et~al.}(2015)\citenamefont {Pastor-Satorras}, \citenamefont {Castellano}, \citenamefont {Van~Mieghem},\ and\ \citenamefont {Vespignani}}]{pastor2015epidemic}%
  \BibitemOpen
  \bibfield  {author} {\bibinfo {author} {\bibfnamefont {R.}~\bibnamefont {Pastor-Satorras}}, \bibinfo {author} {\bibfnamefont {C.}~\bibnamefont {Castellano}}, \bibinfo {author} {\bibfnamefont {P.}~\bibnamefont {Van~Mieghem}},\ and\ \bibinfo {author} {\bibfnamefont {A.}~\bibnamefont {Vespignani}},\ }\bibfield  {title} {\bibinfo {title} {Epidemic processes in complex networks},\ }\href@noop {} {\bibfield  {journal} {\bibinfo  {journal} {Reviews of modern physics}\ }\textbf {\bibinfo {volume} {87}},\ \bibinfo {pages} {925} (\bibinfo {year} {2015})}\BibitemShut {NoStop}%
\bibitem [{\citenamefont {Barabási}\ and\ \citenamefont {Pósfai}(2016)}]{barabasi2016network}%
  \BibitemOpen
  \bibfield  {author} {\bibinfo {author} {\bibfnamefont {A.-L.}\ \bibnamefont {Barabási}}\ and\ \bibinfo {author} {\bibfnamefont {M.}~\bibnamefont {Pósfai}},\ }\href {http://barabasi.com/networksciencebook/} {\emph {\bibinfo {title} {Network science}}}\ (\bibinfo  {publisher} {Cambridge University Press},\ \bibinfo {address} {Cambridge},\ \bibinfo {year} {2016})\BibitemShut {NoStop}%
\bibitem [{\citenamefont {Newman}(2018)}]{newman2018networks}%
  \BibitemOpen
  \bibfield  {author} {\bibinfo {author} {\bibfnamefont {M.}~\bibnamefont {Newman}},\ }\href@noop {} {\emph {\bibinfo {title} {Networks}}}\ (\bibinfo  {publisher} {Oxford university press},\ \bibinfo {year} {2018})\BibitemShut {NoStop}%
\bibitem [{\citenamefont {Zhao}\ \emph {et~al.}(2013)\citenamefont {Zhao}, \citenamefont {Cui}, \citenamefont {Qiu}, \citenamefont {Wang},\ and\ \citenamefont {Wang}}]{zhao2013sir}%
  \BibitemOpen
  \bibfield  {author} {\bibinfo {author} {\bibfnamefont {L.}~\bibnamefont {Zhao}}, \bibinfo {author} {\bibfnamefont {H.}~\bibnamefont {Cui}}, \bibinfo {author} {\bibfnamefont {X.}~\bibnamefont {Qiu}}, \bibinfo {author} {\bibfnamefont {X.}~\bibnamefont {Wang}},\ and\ \bibinfo {author} {\bibfnamefont {J.}~\bibnamefont {Wang}},\ }\bibfield  {title} {\bibinfo {title} {{SIR rumor spreading model in the new media age}},\ }\href@noop {} {\bibfield  {journal} {\bibinfo  {journal} {Physica A: Statistical Mechanics and its Applications}\ }\textbf {\bibinfo {volume} {392}},\ \bibinfo {pages} {995} (\bibinfo {year} {2013})}\BibitemShut {NoStop}%
\bibitem [{\citenamefont {Cinelli}\ \emph {et~al.}(2021)\citenamefont {Cinelli}, \citenamefont {De~Francisci~Morales}, \citenamefont {Galeazzi}, \citenamefont {Quattrociocchi},\ and\ \citenamefont {Starnini}}]{cinelli2021echo}%
  \BibitemOpen
  \bibfield  {author} {\bibinfo {author} {\bibfnamefont {M.}~\bibnamefont {Cinelli}}, \bibinfo {author} {\bibfnamefont {G.}~\bibnamefont {De~Francisci~Morales}}, \bibinfo {author} {\bibfnamefont {A.}~\bibnamefont {Galeazzi}}, \bibinfo {author} {\bibfnamefont {W.}~\bibnamefont {Quattrociocchi}},\ and\ \bibinfo {author} {\bibfnamefont {M.}~\bibnamefont {Starnini}},\ }\bibfield  {title} {\bibinfo {title} {The echo chamber effect on social media},\ }\href@noop {} {\bibfield  {journal} {\bibinfo  {journal} {Proceedings of the National Academy of Sciences}\ }\textbf {\bibinfo {volume} {118}},\ \bibinfo {pages} {e2023301118} (\bibinfo {year} {2021})}\BibitemShut {NoStop}%
\bibitem [{\citenamefont {Granell}\ \emph {et~al.}(2013)\citenamefont {Granell}, \citenamefont {G{\'o}mez},\ and\ \citenamefont {Arenas}}]{granell2013dynamical}%
  \BibitemOpen
  \bibfield  {author} {\bibinfo {author} {\bibfnamefont {C.}~\bibnamefont {Granell}}, \bibinfo {author} {\bibfnamefont {S.}~\bibnamefont {G{\'o}mez}},\ and\ \bibinfo {author} {\bibfnamefont {A.}~\bibnamefont {Arenas}},\ }\bibfield  {title} {\bibinfo {title} {Dynamical interplay between awareness and epidemic spreading in multiplex networks},\ }\href@noop {} {\bibfield  {journal} {\bibinfo  {journal} {Physical review letters}\ }\textbf {\bibinfo {volume} {111}},\ \bibinfo {pages} {128701} (\bibinfo {year} {2013})}\BibitemShut {NoStop}%
\bibitem [{\citenamefont {Granell}\ \emph {et~al.}(2014)\citenamefont {Granell}, \citenamefont {G{\'o}mez},\ and\ \citenamefont {Arenas}}]{granell2014competing}%
  \BibitemOpen
  \bibfield  {author} {\bibinfo {author} {\bibfnamefont {C.}~\bibnamefont {Granell}}, \bibinfo {author} {\bibfnamefont {S.}~\bibnamefont {G{\'o}mez}},\ and\ \bibinfo {author} {\bibfnamefont {A.}~\bibnamefont {Arenas}},\ }\bibfield  {title} {\bibinfo {title} {Competing spreading processes on multiplex networks: awareness and epidemics},\ }\href@noop {} {\bibfield  {journal} {\bibinfo  {journal} {Physical review E}\ }\textbf {\bibinfo {volume} {90}},\ \bibinfo {pages} {012808} (\bibinfo {year} {2014})}\BibitemShut {NoStop}%
\bibitem [{\citenamefont {Guo}\ \emph {et~al.}(2015)\citenamefont {Guo}, \citenamefont {Jiang}, \citenamefont {Lei}, \citenamefont {Li}, \citenamefont {Ma},\ and\ \citenamefont {Zheng}}]{guo2015two}%
  \BibitemOpen
  \bibfield  {author} {\bibinfo {author} {\bibfnamefont {Q.}~\bibnamefont {Guo}}, \bibinfo {author} {\bibfnamefont {X.}~\bibnamefont {Jiang}}, \bibinfo {author} {\bibfnamefont {Y.}~\bibnamefont {Lei}}, \bibinfo {author} {\bibfnamefont {M.}~\bibnamefont {Li}}, \bibinfo {author} {\bibfnamefont {Y.}~\bibnamefont {Ma}},\ and\ \bibinfo {author} {\bibfnamefont {Z.}~\bibnamefont {Zheng}},\ }\bibfield  {title} {\bibinfo {title} {Two-stage effects of awareness cascade on epidemic spreading in multiplex networks},\ }\href@noop {} {\bibfield  {journal} {\bibinfo  {journal} {Physical Review E}\ }\textbf {\bibinfo {volume} {91}},\ \bibinfo {pages} {012822} (\bibinfo {year} {2015})}\BibitemShut {NoStop}%
\bibitem [{\citenamefont {Guo}\ \emph {et~al.}(2016)\citenamefont {Guo}, \citenamefont {Lei}, \citenamefont {Jiang}, \citenamefont {Ma}, \citenamefont {Huo},\ and\ \citenamefont {Zheng}}]{guo2016epidemic}%
  \BibitemOpen
  \bibfield  {author} {\bibinfo {author} {\bibfnamefont {Q.}~\bibnamefont {Guo}}, \bibinfo {author} {\bibfnamefont {Y.}~\bibnamefont {Lei}}, \bibinfo {author} {\bibfnamefont {X.}~\bibnamefont {Jiang}}, \bibinfo {author} {\bibfnamefont {Y.}~\bibnamefont {Ma}}, \bibinfo {author} {\bibfnamefont {G.}~\bibnamefont {Huo}},\ and\ \bibinfo {author} {\bibfnamefont {Z.}~\bibnamefont {Zheng}},\ }\bibfield  {title} {\bibinfo {title} {Epidemic spreading with activity-driven awareness diffusion on multiplex network},\ }\href@noop {} {\bibfield  {journal} {\bibinfo  {journal} {Chaos: An Interdisciplinary Journal of Nonlinear Science}\ }\textbf {\bibinfo {volume} {26}} (\bibinfo {year} {2016})}\BibitemShut {NoStop}%
\bibitem [{\citenamefont {Peng}\ \emph {et~al.}(2021)\citenamefont {Peng}, \citenamefont {Lu}, \citenamefont {Lin}, \citenamefont {Lindstrom}, \citenamefont {Parkinson}, \citenamefont {Wang}, \citenamefont {Bertozzi},\ and\ \citenamefont {Porter}}]{peng2021multilayer}%
  \BibitemOpen
  \bibfield  {author} {\bibinfo {author} {\bibfnamefont {K.}~\bibnamefont {Peng}}, \bibinfo {author} {\bibfnamefont {Z.}~\bibnamefont {Lu}}, \bibinfo {author} {\bibfnamefont {V.}~\bibnamefont {Lin}}, \bibinfo {author} {\bibfnamefont {M.~R.}\ \bibnamefont {Lindstrom}}, \bibinfo {author} {\bibfnamefont {C.}~\bibnamefont {Parkinson}}, \bibinfo {author} {\bibfnamefont {C.}~\bibnamefont {Wang}}, \bibinfo {author} {\bibfnamefont {A.~L.}\ \bibnamefont {Bertozzi}},\ and\ \bibinfo {author} {\bibfnamefont {M.~A.}\ \bibnamefont {Porter}},\ }\bibfield  {title} {\bibinfo {title} {A multilayer network model of the coevolution of the spread of a disease and competing opinions},\ }\href@noop {} {\bibfield  {journal} {\bibinfo  {journal} {Mathematical Models and Methods in Applied Sciences}\ }\textbf {\bibinfo {volume} {31}},\ \bibinfo {pages} {2455} (\bibinfo {year} {2021})}\BibitemShut {NoStop}%
\bibitem [{\citenamefont {Vel\'asquez-Rojas}\ \emph {et~al.}(2020)\citenamefont {Vel\'asquez-Rojas}, \citenamefont {Ventura}, \citenamefont {Connaughton}, \citenamefont {Moreno}, \citenamefont {Rodrigues},\ and\ \citenamefont {Vazquez}}]{PhysRevE.102.022312}%
  \BibitemOpen
  \bibfield  {author} {\bibinfo {author} {\bibfnamefont {F.}~\bibnamefont {Vel\'asquez-Rojas}}, \bibinfo {author} {\bibfnamefont {P.~C.}\ \bibnamefont {Ventura}}, \bibinfo {author} {\bibfnamefont {C.}~\bibnamefont {Connaughton}}, \bibinfo {author} {\bibfnamefont {Y.}~\bibnamefont {Moreno}}, \bibinfo {author} {\bibfnamefont {F.~A.}\ \bibnamefont {Rodrigues}},\ and\ \bibinfo {author} {\bibfnamefont {F.}~\bibnamefont {Vazquez}},\ }\bibfield  {title} {\bibinfo {title} {Disease and information spreading at different speeds in multiplex networks},\ }\href {https://doi.org/10.1103/PhysRevE.102.022312} {\bibfield  {journal} {\bibinfo  {journal} {Phys. Rev. E}\ }\textbf {\bibinfo {volume} {102}},\ \bibinfo {pages} {022312} (\bibinfo {year} {2020})}\BibitemShut {NoStop}%
\bibitem [{\citenamefont {Xia}\ \emph {et~al.}(2019)\citenamefont {Xia}, \citenamefont {Wang}, \citenamefont {Zheng}, \citenamefont {Guo}, \citenamefont {Shi}, \citenamefont {Dehmer},\ and\ \citenamefont {Chen}}]{xia2019new}%
  \BibitemOpen
  \bibfield  {author} {\bibinfo {author} {\bibfnamefont {C.}~\bibnamefont {Xia}}, \bibinfo {author} {\bibfnamefont {Z.}~\bibnamefont {Wang}}, \bibinfo {author} {\bibfnamefont {C.}~\bibnamefont {Zheng}}, \bibinfo {author} {\bibfnamefont {Q.}~\bibnamefont {Guo}}, \bibinfo {author} {\bibfnamefont {Y.}~\bibnamefont {Shi}}, \bibinfo {author} {\bibfnamefont {M.}~\bibnamefont {Dehmer}},\ and\ \bibinfo {author} {\bibfnamefont {Z.}~\bibnamefont {Chen}},\ }\bibfield  {title} {\bibinfo {title} {A new coupled disease-awareness spreading model with mass media on multiplex networks},\ }\href@noop {} {\bibfield  {journal} {\bibinfo  {journal} {Information Sciences}\ }\textbf {\bibinfo {volume} {471}},\ \bibinfo {pages} {185} (\bibinfo {year} {2019})}\BibitemShut {NoStop}%
\bibitem [{\citenamefont {Fang}\ \emph {et~al.}(2023)\citenamefont {Fang}, \citenamefont {Ma},\ and\ \citenamefont {Li}}]{fang2023coevolution}%
  \BibitemOpen
  \bibfield  {author} {\bibinfo {author} {\bibfnamefont {F.}~\bibnamefont {Fang}}, \bibinfo {author} {\bibfnamefont {J.}~\bibnamefont {Ma}},\ and\ \bibinfo {author} {\bibfnamefont {Y.}~\bibnamefont {Li}},\ }\bibfield  {title} {\bibinfo {title} {The coevolution of the spread of a disease and competing opinions in multiplex networks},\ }\href@noop {} {\bibfield  {journal} {\bibinfo  {journal} {Chaos, Solitons \& Fractals}\ }\textbf {\bibinfo {volume} {170}},\ \bibinfo {pages} {113376} (\bibinfo {year} {2023})}\BibitemShut {NoStop}%
\bibitem [{\citenamefont {Dubois}\ and\ \citenamefont {Blank}(2018)}]{dubois2018echo}%
  \BibitemOpen
  \bibfield  {author} {\bibinfo {author} {\bibfnamefont {E.}~\bibnamefont {Dubois}}\ and\ \bibinfo {author} {\bibfnamefont {G.}~\bibnamefont {Blank}},\ }\bibfield  {title} {\bibinfo {title} {The echo chamber is overstated: the moderating effect of political interest and diverse media},\ }\href@noop {} {\bibfield  {journal} {\bibinfo  {journal} {Information, communication \& society}\ }\textbf {\bibinfo {volume} {21}},\ \bibinfo {pages} {729} (\bibinfo {year} {2018})}\BibitemShut {NoStop}%
\bibitem [{\citenamefont {Mason}(2015)}]{mason2015disrespectfully}%
  \BibitemOpen
  \bibfield  {author} {\bibinfo {author} {\bibfnamefont {L.}~\bibnamefont {Mason}},\ }\bibfield  {title} {\bibinfo {title} {{“I disrespectfully agree”: The differential effects of partisan sorting on social and issue polarization}},\ }\href@noop {} {\bibfield  {journal} {\bibinfo  {journal} {American journal of political science}\ }\textbf {\bibinfo {volume} {59}},\ \bibinfo {pages} {128} (\bibinfo {year} {2015})}\BibitemShut {NoStop}%
\bibitem [{\citenamefont {Axelrod}(1997)}]{axelrod1997dissemination}%
  \BibitemOpen
  \bibfield  {author} {\bibinfo {author} {\bibfnamefont {R.}~\bibnamefont {Axelrod}},\ }\bibfield  {title} {\bibinfo {title} {The dissemination of culture: A model with local convergence and global polarization},\ }\href@noop {} {\bibfield  {journal} {\bibinfo  {journal} {Journal of conflict resolution}\ }\textbf {\bibinfo {volume} {41}},\ \bibinfo {pages} {203} (\bibinfo {year} {1997})}\BibitemShut {NoStop}%
\bibitem [{\citenamefont {Holme}\ and\ \citenamefont {Kim}(2002)}]{holme2002growing}%
  \BibitemOpen
  \bibfield  {author} {\bibinfo {author} {\bibfnamefont {P.}~\bibnamefont {Holme}}\ and\ \bibinfo {author} {\bibfnamefont {B.~J.}\ \bibnamefont {Kim}},\ }\bibfield  {title} {\bibinfo {title} {Growing scale-free networks with tunable clustering},\ }\href@noop {} {\bibfield  {journal} {\bibinfo  {journal} {Physical review E}\ }\textbf {\bibinfo {volume} {65}},\ \bibinfo {pages} {026107} (\bibinfo {year} {2002})}\BibitemShut {NoStop}%
\bibitem [{\citenamefont {Albert}\ and\ \citenamefont {Barab{\'a}si}(2002)}]{albert2002statistical}%
  \BibitemOpen
  \bibfield  {author} {\bibinfo {author} {\bibfnamefont {R.}~\bibnamefont {Albert}}\ and\ \bibinfo {author} {\bibfnamefont {A.-L.}\ \bibnamefont {Barab{\'a}si}},\ }\bibfield  {title} {\bibinfo {title} {Statistical mechanics of complex networks},\ }\href@noop {} {\bibfield  {journal} {\bibinfo  {journal} {Reviews of modern physics}\ }\textbf {\bibinfo {volume} {74}},\ \bibinfo {pages} {47} (\bibinfo {year} {2002})}\BibitemShut {NoStop}%
\bibitem [{\citenamefont {Fine}\ \emph {et~al.}(2011)\citenamefont {Fine}, \citenamefont {Eames},\ and\ \citenamefont {Heymann}}]{fine2011herd}%
  \BibitemOpen
  \bibfield  {author} {\bibinfo {author} {\bibfnamefont {P.}~\bibnamefont {Fine}}, \bibinfo {author} {\bibfnamefont {K.}~\bibnamefont {Eames}},\ and\ \bibinfo {author} {\bibfnamefont {D.~L.}\ \bibnamefont {Heymann}},\ }\bibfield  {title} {\bibinfo {title} {“herd immunity”: a rough guide},\ }\href@noop {} {\bibfield  {journal} {\bibinfo  {journal} {Clinical infectious diseases}\ }\textbf {\bibinfo {volume} {52}},\ \bibinfo {pages} {911} (\bibinfo {year} {2011})}\BibitemShut {NoStop}%
\bibitem [{\citenamefont {Komuro}(2025)}]{git_repo}%
  \BibitemOpen
  \bibfield  {author} {\bibinfo {author} {\bibfnamefont {S.}~\bibnamefont {Komuro}},\ }\href@noop {} {\bibinfo {title} {{Modelling the impact of polarisation on epidemics}}},\ \bibinfo {howpublished} {\url{https://github.com/skomuro/polarisation_meets_epi}} (\bibinfo {year} {2025}),\ \bibinfo {note} {accessed: 2025-01-20}\BibitemShut {NoStop}%
\end{thebibliography}%

\end{document}